\pdfoutput=1

\documentclass[a4paper,twocolumn,showpacs,showkeys,amsmath,amssymb,aps,pra,]{revtex4-1}

\usepackage{amsfonts}
\usepackage{dsfont}
\usepackage{graphicx}
\usepackage{hyperref}
\usepackage{subfigure}

\newcommand{\ketbra}[2]{| #1 \rangle \langle #2 |}
\newcommand{\ket}[1]{| #1 \rangle}
\newcommand{\bra}[1]{\langle #1 |}

\newcommand{\tr}{\mathrm{tr}\,}

\newcommand{\ii}{\mathrm{i}}

\newcommand{\ie}{\emph{i.e.}}
\newcommand{\eg}{\emph{e.g.}}

\newcommand{\Expectation}{\ensuremath{\mathrm{E}}}

\begin{document}
\title{Probability measure generated by the superfidelity}
\author{Zbigniew Pucha{\l}a}
\email{z.puchala@iitis.pl}
\author{Jaros{\l}aw Adam Miszczak}
\email{miszczak@iitis.pl}
\affiliation{Institute of Theoretical and Applied Informatics, Polish Academy
of Sciences, Ba{\l}tycka 5, 44-100 Gliwice, Poland}

\begin{abstract}
We study the probability measure on the space of density matrices induced by the
metric defined by using superfidelity. We give the formula for the probability
density of eigenvalues. We also study some statistical properties of the set of
density matrices equipped with the introduced measure and provide a method
for generating density matrices according to the introduced measure.
\end{abstract}

\date{14/07/2011 (v. 2.04)}

\maketitle

\section{Introduction}
Recent applications of quantum mechanics are based on processing and
transferring information encoded in quantum states. Random quantum states can be
used to study various effects unique to quantum information
theory~\cite{zyczkowski11generating}. This is especially true if one needs to
get some information about the typical properties of the system in
question~\cite{BZ06}. In many cases it is important to quantify to what degree
states are similar to the average state or how, on average, given quantity
evolves during the execution of a quantum procedure. The crucial question
emerging in this situation is how one should choose random sample from the set
of quantum states.

The aforementioned question be answered easily in the case of pure quantum
states. In this situation there exists a single, natural measure for
constructing ensembles of states, namely Fubini-Study measure. The situation is
more complex in the case of mixed quantum states. The probability measure can be
introduced using various distance measures between quantum states~\cite{BZ06}.
By choosing the metric we also choose the probability measure on the space of
density matrices. Among the most commonly used metrics we can point out the
trace distance, Hilbert-Schmidt distance, and Bures distance. 

In the analysis of mixed quantum states Bures distance is the most commonly used
metric among the ones mentioned above. It has many important
properties~\cite{BZ06}. In particular it is a Riemannian and monotone metric. On
the space of pure states it reduces to Fubini-Study metric~\cite{Uh76} and it
induces the statistical distance in the subspace of diagonal density
matrices~\cite{caves}.

The main aim of this paper is the analysis of the probability measure on the
space of density matrices induced by the metric defined in terms of
\emph{superfidelity}~\cite{subsuper}. We calculate the formula for the
probability density of eigenvalues and study some properties of the space of
quantum states equipped with the introduced measure. We also provide a method
for sampling random density matrices according to the introduced distribution.

This paper is organized as follows.
In Section \ref{sec:intro} we introduce notation and basic facts used in the
following sections.
In Section~\ref{sec:probmeasure} we calculate the volume element for the measure
generated by the metric based on superfidelity and compare it with the analogous
metric based on quantum fidelity.
In Section~\ref{sec:density} we provide a formula for a probability density
function on a simplex of eigenvalues. We also calculate the normalization
constant in the low-dimensional case.
In Section~\ref{sec:generating} we provide a method for sampling density
matrices according to the introduced measure.
Finally, in Section \ref{sec:summary} we provide a summary of the presented
results.

\section{Preliminaries}\label{sec:intro}
Let use denote by $\mathcal{M}_N$ the space of density matrices of size $N$,
\ie\ $N\times N$ positive matrices with unit trace. By $\Delta$ we denote the
simplex of eigenvalues. 

For two density matrices $\rho,\sigma\in\mathcal{M}_N$, Bures distance can be
defined in terms of quantum fidelity~\cite{Uh76} as
\begin{equation}\label{eqn:def-bures}
d_\mathrm{B}(\rho,\sigma) = \sqrt{2-2\sqrt{F(\rho,\sigma)}},
\end{equation}
where fidelity,  
$
F(\rho,\sigma)= \left[\tr|\sqrt{\rho}\sqrt{\sigma}|\right]^2,
$
provides a measure of similarity on the space of density matrices.

Probability measure on the simplex of eigenvalues generated by Bures metric
was calculated in~\cite{Hubner92, Hall98, buresvolume}. Various statistical
properties of ensembles of quantum states with respect to this measure were
discussed in \cite{sommers04statistical}.

Bures distance is commonly used in quantum information theory as a natural
metric on the space of density matrices. Unfortunately, fidelity used to express
$d_\mathrm{B}$ has some serious drawbacks. In particular in order to calculate
fidelity between two quantum states one needs to compute square root of matrix,
which is in general computationally hard task. Also, fidelity cannot be measured
directly in laboratory and thus cannot be used to analyse experiments directly.

These drawbacks motivated the introduction of a new measure of similarity,
namely superfidelity~\cite{subsuper}, defined for $\rho,\sigma\in\mathcal{M}_N$
as
\begin{equation}
G(\rho,\sigma)=\tr\rho\sigma + \sqrt{1-\tr\rho^2}\sqrt{1-\tr\sigma^2}.
\end{equation}
Superfidelity shares many features with fidelity, \ie\ it is bounded, symmetric
and unitarly invariant. Moreover it is jointly concave and supermultiplicative.
It was proved that superfidelity gives an upper bound for fidelity,
$F(\rho,\sigma) \leq G(\rho,\sigma)$, where the equality is for $\rho, \sigma
\in \mathcal{M}_2$ or in the case where one of the states is pure. It was also
shown that, although $G$ is not monotone~\cite{mendonca}, it can be used to
define metric on $\mathcal{M}_N$.
Using the correspondence between quantum operations and quantum states,
superfidelity can be used to introduce metric on the space of quantum
channels~\cite{process-superfidelity}. Superfidelity was also proved to be
useful in providing bounds on the trace distance~\cite{boundtrace} (\ie\
distinguisabiliy of states~\cite{englert96fringe}) and as a tool for studying
new metrics on the space of quantum states~\cite{chen2010super}.

In the following we use a metric on the space of density matrices defined for
$\rho,\sigma\in\mathcal{M}_N$~as
\begin{equation}\label{eqn:supermetric}
d_G(\rho,\sigma) = \sqrt{2 - 2 G(\rho,\sigma)}.
\end{equation}
Before we discuss further properties of this metric we should stress that the
direct analogous of the Bures distance,
$
d_{\sqrt{G}}(\rho,\sigma)=\sqrt{2-2\sqrt{G(\rho,\sigma)}},
$ is not a metric.
One should also note that, since $G$ is not monotone it cannot be analysed using
Morozova-\v{C}encov-Petz theorem~\cite{BZ06}. 

\section{Volume element for the measure}\label{sec:probmeasure}
To obtain the probability measure induced by metric Eq.~(\ref{eqn:supermetric})
one needs to derive the volume element.

The calculations below follow the approach used by H{\"u}bner~\cite{Hubner92}.
We begin with the calculation of the line element
\begin{equation}
d_G^2(\rho, \rho + d \rho) = 2 -2 G(\rho,\rho + d \rho).
\end{equation}
We introduce function 
$
A(t) = G(\rho,\rho +  t d \rho),
$
which allows to write the line element
\begin{equation}
g_{ij} d\rho^i d\rho^j = \frac{1}{2} 
 \frac{d^2}{d t^2}[d_G^2(\rho, \rho + t \; d \rho)]  \Big\vert_{t=0}
\end{equation}
as 
\begin{equation}
g_{ij} d\rho^i d\rho^j = - A''(t) \Big\vert_{t=0}.
\end{equation}
Equivalently, with the use of matrix entries, the line element reads
\begin{equation}\label{eqn:s2super}
g_{ij} d\rho^i d\rho^j = \frac{(\sum \lambda_i \bra{i} d \rho \ket{i})^2}
                               {1 - \sum \lambda_i^2} 
                       + \sum \bra{i} (d \rho)^2 \ket{i}.
\end{equation}

Infinitesimal shift $\rho + d \rho$ can be decomposed as a shift in eigenvalues
and infinitesimal unitary rotation~\cite{Hall98}
\begin{equation}
\rho + d \rho = \rho + d \Lambda + [d U, \rho],
\end{equation}
where $d \Lambda = \sum d \lambda_i \ketbra{i}{i}$ and $(dU)^{\dagger} = - dU$.
Rewriting $dU$ in computational basis gives 
\begin{equation}
dU = \sum_{j,k} (dx_{jk} + \ii d y_{jk} ) \ketbra{j}{k}
\end{equation}
with real coefficients $dx_{jk} = - dx_{kj}$ and $dy_{jk} = dy_{kj}$.
After some calculations one gets
\begin{equation}
 \tr d \rho^2 = \sum_{i} (d \lambda_i)^2 + 
 2 \sum_{i < j} (\lambda_i - \lambda_j)^2 [(d x_{ij})^2 + (d y_{ij})^2]
\end{equation}
and
\begin{equation}
 \tr \rho d \rho = \sum_{i} \lambda_i d\lambda_i.
\end{equation}
Expanding this we get the entries of the metric tensor
\begin{equation}
\begin{split}
g_{ij} d\rho^i d\rho^j & = 
\sum_{i,j} \left(\frac{\lambda_i \lambda_j}{1-\tr \rho^2} + \delta_{ij} \right) 
   d\lambda_i d\lambda_j 
 \\ 
 & + 2 \sum_{i < j} (\lambda_i - \lambda_j)^2 [(d x_{ij})^2 + (d y_{ij})^2].
\end{split}
\end{equation}

To obtain volume element of the sought measure, one must calculate the 
appropriate determinant
\begin{equation}
\begin{split}
dV_G & = \sqrt{\det{\left(  \frac{ \lambda_i \lambda_j }{1-\tr \rho^2} 
         + \delta_{ij} \right)}} d\lambda_1 \dots d\lambda_n  
     \\
     &\times \prod_{i < j} 2 (\lambda_i - \lambda_j)^2 d x_{ij}d y_{ij}.
\end{split}
\end{equation}
Using the equality
\begin{equation}
\det{\left(  \frac{ \lambda_i \lambda_j }{1-\tr \rho^2} + \delta_{ij} \right)}  
 = 1 + \frac{\tr \rho^2}{1-\tr \rho^2} = \frac{1}{1-\tr \rho^2},
\end{equation}
we obtain the expression for the volume element
\begin{equation}\label{eqn:super-volume-element}
dV_G =  \frac{d\lambda_1 \dots d\lambda_n}{\sqrt{1-\sum_{i} \lambda_i^2}}  
        \prod_{i < j} 2 (\lambda_i - \lambda_j)^2 d x_{ij}d y_{ij}.
\end{equation}

One can compare the above formulas for the line element with the analogous
result for the metric given in terms of fidelity as
\begin{equation}
d_{B'}^2(\rho,\rho+d \rho) = 2\left(1-F(\rho,\rho+d \rho)\right).
\end{equation}
In this case it is easy to check that the line element is given by formula
\begin{equation}
d^2_{B'}(\rho,\rho+d \rho) 
=\sum_{ij}\frac{|\bra{i}d\rho \ket{j}|^2}{\lambda_i+\lambda_j} .
\end{equation}
In the one-qubit case the above formula reads
\begin{equation}
d^2_{B'}(\rho,\rho+d \rho) =
\left( \frac{1}{2 \lambda (1 - \lambda)} \right) |d\rho_{11}|^2
+ |d\rho_{12}|^2 + |d\rho_{21}|^2
\end{equation}
where $\lambda$ and $1-\lambda$ are eigenvalues of $\rho$ and $d\rho_{ij}=
\bra{i}d\rho \ket{j} $ and we have used the equality $\bra{1}d\rho \ket{1} = -
\bra{2}d\rho \ket{2} $. This is identical to (\ref{eqn:s2super}) for $N=2$,
which is what one expects since in this case $F(\rho,\sigma) =
G(\rho,\sigma)$.

\section{Probability density function}\label{sec:density}
In order to obtain probability measure we need to specify the normalizing
constant. This constant is an inverse of the integral of the volume element
$dV_G$ over the group of unitary matrices and over the simplex of eigenvalues.

\subsection{Normalization constant}
Integration with respect to $U(N)$ is independent from the integration over the 
simplex of eigenvalues. We can rewrite Eq.~\ref{eqn:super-volume-element} as
\begin{equation}
dV_G =  \left( \frac{2^{N(N-1)/2}}{\sqrt{1-\sum_{i} \lambda_i^2}} 
   \prod_{i < j}  (\lambda_i - \lambda_j)^2 \right) d\lambda_1 \dots d\lambda_n
   \prod_{i \neq j}  d x_{ij}d y_{ij}.
\end{equation}
After integrating this formula over $U(N)$ we get
\begin{equation}
V_G =  \Upsilon_N 
       \int_{\Delta}  \left( \frac{2^{N(N-1)/2}}{\sqrt{1-\sum_{i} \lambda_i^2}}  
       \prod_{i < j}  (\lambda_i - \lambda_j)^2 \right) 
       d\lambda_1 \dots d\lambda_n,
\end{equation}
where $\Upsilon_N $ is the volume of projective $U(N)$ \cite[Eq.~(148)]{caves}
\begin{equation}
\Upsilon_N  = \frac{ \pi^{N(N-1)/2}}{\prod_{d=1}^{N-1} d !}
\end{equation}
and $\Delta$ is the simplex of eigenvalues.

\begin{figure}[h!]
	\centering
	\subfigure[Measure generated by
	$\sqrt{1-G}$ metric.]{\includegraphics[width=0.45\columnwidth]
	{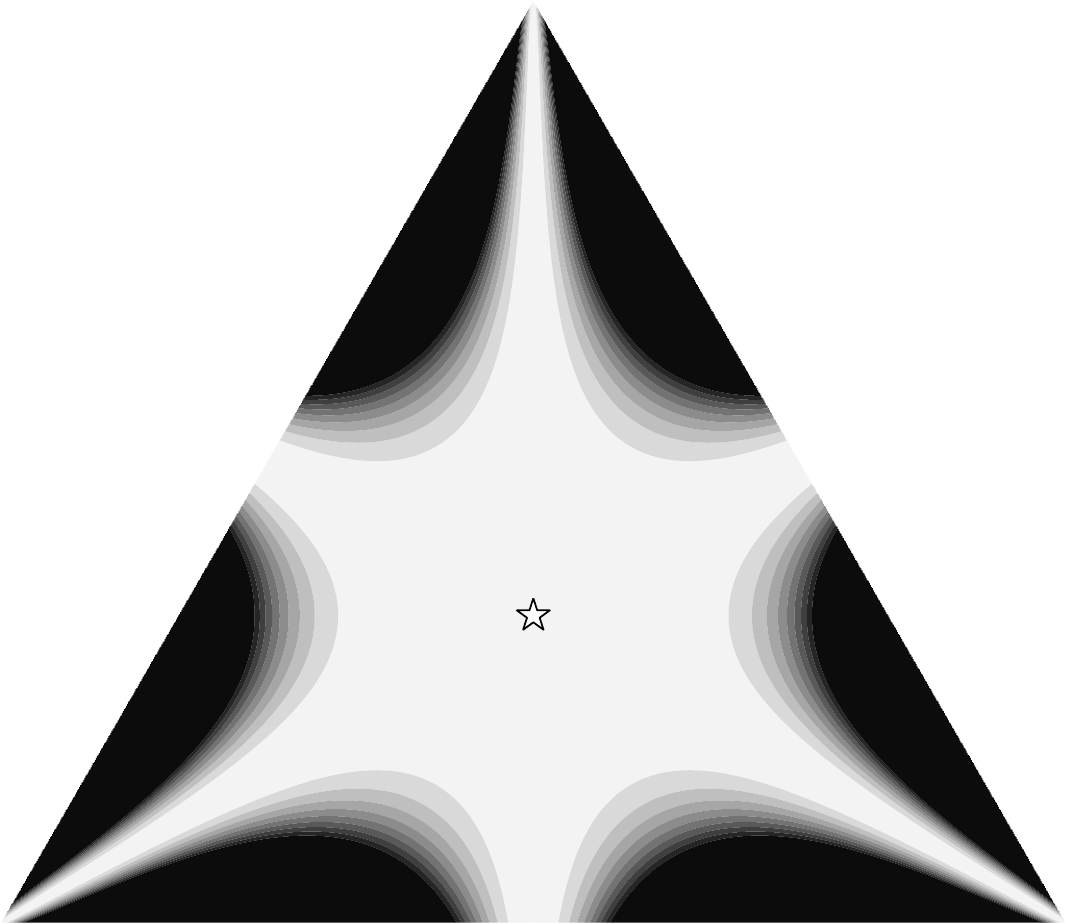}\label{fig:super-density-3}}\quad 
	\subfigure[Measure generated by the
	Bures metric]{\includegraphics[width=0.45\columnwidth]
	{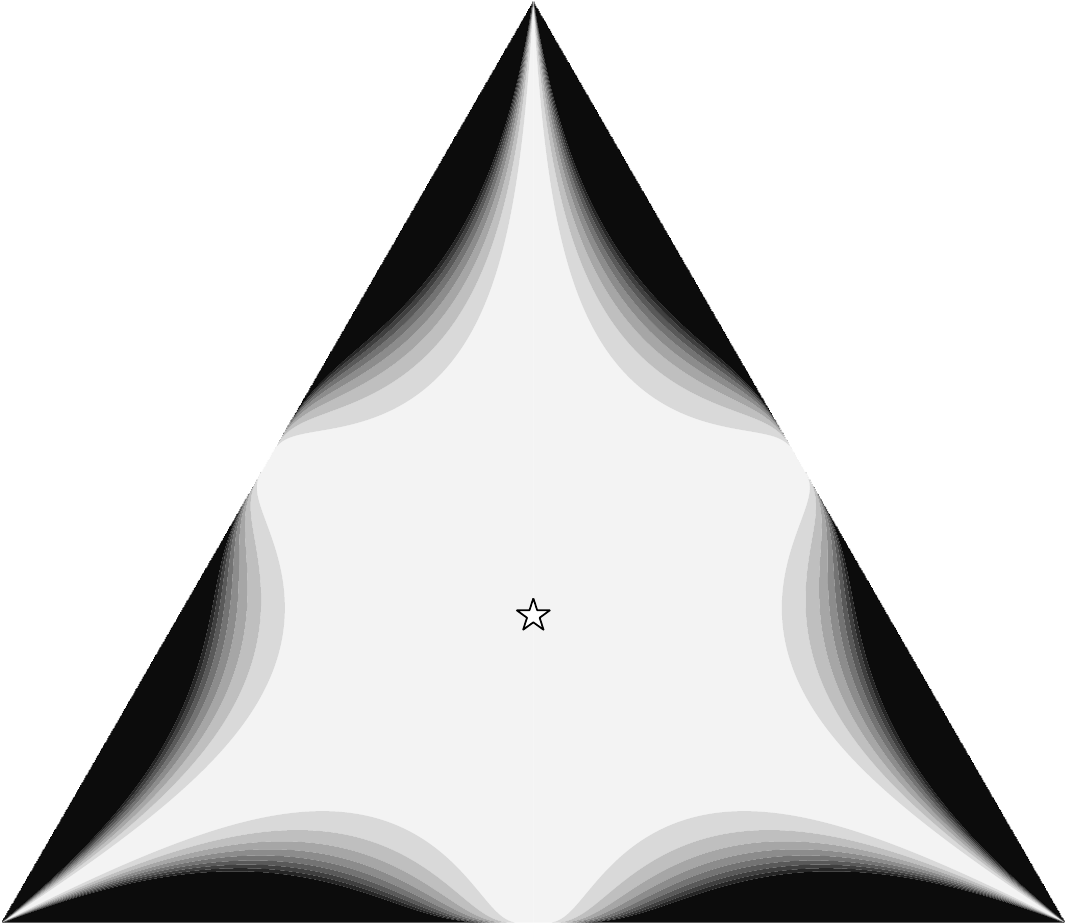}\label{fig:bures-density-3}}
	\caption{Distribution of the eigenvalues for one-qutrit ($N=3$) density
	matrices for different probability measures.}
	\label{fig:density-3}
\end{figure}

Probability density function on a simplex of eigenvalues is given by 
\begin{equation} \label{eqn:pdf-GN}
f_{\mathrm{G},N} (\lambda) =  C^{\mathrm{G}}_N \prod_{i < j}  (\lambda_i - \lambda_j)^2  \frac{1}{\sqrt{1 - \sum_i \lambda_i^2}},
\end{equation}
where $C_N$ is a normalization constant. For $N=3$ function $f_{\mathrm{G},N} $
is presented in~Fig.~\ref{fig:super-density-3}.

The normalization constant $C_N^\mathrm{G}$ is  the following integral
\begin{equation}\label{eqn:calka}
\frac{1}{C_N^\mathrm{G}} = \int_{\Delta}       
\prod_{i < j}  (\lambda_i - \lambda_j)^2  \frac{1}{\sqrt{1 - \sum_i \lambda_i^2}}
d\lambda
\end{equation}
over the simplex of eigenvalues.

The above integral can be written in terms of expectation value with respect to 
Hilbert-Schmidt measure
\begin{equation}\label{eqn:cgn-definition}
\frac{1}{C^\mathrm{G}_N}  = \frac{1}{C^{\mathrm{HS}}_N}\Expectation \left[\frac{1}{\sqrt{1 - \tr \rho^2}} \right],
\end{equation}
where $\rho$ is a random state distributed with Hilbert-Schmidt measure and
\begin{equation}
C_N^{\mathrm{HS}}=\frac{\Gamma(N^2)}{\prod_{k=1}^N\Gamma(k)\Gamma(k+1)}.
\end{equation}

The distribution of purity $(\tr \rho^2)$ for random states is a subject of much
study~\cite{giraud_distribution_2007, giraud2007purity, scott2003entangling}.

The probability distribution function of purity is known for Hilbert-Schmidt 
random states in the case of $N = 2$ and $N=3$ \cite{giraud2007purity}. Using 
these results, we can write explicitly normalizing constants
\begin{eqnarray}
	C_2^\mathrm{G}&=& \frac{2 \sqrt{2}}{3 \pi } C^{\mathrm{HS}}_2,\\
	C_3^\mathrm{G}&=& \frac{432 \sqrt{2}}{317 \pi } C^{\mathrm{HS}}_3. 
\end{eqnarray}

In the case of $N>3$ one can use the series expansion of $\frac{1}{\sqrt{1-r}}$
and rewrite the above as 
\begin{eqnarray}
\frac{1}{C_N^\mathrm{G}} &=&       
   \frac{1}{C^{\mathrm{HS}}_N} \sum_{k=0}^{\infty} \frac{(2k-1)!!}{k! 2^k}    
   \Expectation[ \tr(\rho^2)^k].
\end{eqnarray}
The moments of purity for Hilbert-Schmidt random state are given by
\cite{giraud_distribution_2007, giraud2007purity}
\begin{equation}
\begin{split}
\Expectation[(\tr \rho^2)^k] = \frac{N!(N^2 -1)!}{(N^2+2N-1)!}   \sum_{k_1 +  \dots + k_N = k} \frac{k!}{\prod_{i=1}^N k_i! } \times\\ 
 \prod_{i=1}^n \frac{(n+2 k_i - i)!}{(q-i)! i!}   \prod_{1 \leq i < j \leq n} (2 k_i - i - 2 k_j + j).
\end{split}
\end{equation}

The constant $C_N^\mathrm{G}$ can be bounded from the above by using Jensen
inequality
\begin{eqnarray}
 \frac{1}{C_N^\mathrm{G}} &=& \frac{1}{C^{\mathrm{HS}}_N} \Expectation \left[\frac{1}{\sqrt{1 - \tr \rho^2}} \right] \\
 &\geq&  \frac{1}{C^{\mathrm{HS}}_N}  \frac{1}{\sqrt{1 - E [\tr \rho^2]}} = \frac{1}{C^{\mathrm{HS}}_N} \frac{1}{\sqrt{1- \frac{2 N}{N^2+1}}},
\end{eqnarray}
thus
\begin{eqnarray}
 C_N^\mathrm{G} &\leq& C^{\mathrm{HS}}_N \sqrt{1- \frac{2 N}{N^2+1}}.
\end{eqnarray}
Since the distribution of purity has the variance given by
\begin{equation}
 \sigma^2(\tr  \rho^2) = \frac{2 \left(N^2-1\right)^2}{\left(N^2+1\right)^2 \left(N^2+2\right) \left(N^2+3\right)},
\end{equation}
it tends to be more concentrated around the mean given by
\begin{equation}
 \Expectation[\tr\rho^2] =\frac{2N}{N^2+1},
\end{equation}
which tends to zero for large $N$. For small $x$, function $1/\sqrt{1-x}$ can be
approximated with a small error by a linear function. Thus Jensen inequality
gives good approximation of $C_N^\mathrm{G}$ for large values of $N$, where
$\tr\rho^2$ tends to be small.

\subsection{Mean purity}
Let $\rho_G$ be a random state distributed with measure $G$. Then the mean
purity is given as
\begin{equation}
\Expectation[\tr \rho_G^2 ] = \frac{C_N^\mathrm{G}}{C^{\mathrm{HS}}_N} \Expectation \left[\frac{\tr \rho^2}{\sqrt{1 - \tr \rho^2}} \right],
\end{equation}
where $\rho$ has Hilbert-Schmidt distribution. Next we have
\begin{equation}
\Expectation \left[\frac{\tr \rho^2}{\sqrt{1 - \tr \rho^2}} \right] \geq
\Expectation \left[\tr \rho^2\right] \Expectation\left[\frac{1}{\sqrt{1 - \tr \rho^2}} \right],
\end{equation}
which follows from the fact that random variables $\tr \rho^2$ and 
$\frac{1}{\sqrt{1 - \tr \rho^2}}$ are associated (see \eg{}~\cite{tong1997relationship}).
Finally, by using Eq.~(\ref{eqn:cgn-definition}), we get
\begin{equation}
\Expectation    [\tr \rho_G^2 ] 
\geq \frac{C_N^\mathrm{G}}{C^{\mathrm{HS}}_N} \Expectation \left[\tr \rho^2\right] \Expectation\left[\frac{1}{\sqrt{1 - \tr \rho^2}} \right]
= \Expectation \left[\tr \rho^2\right].
\end{equation}

From the above one can see that the mean purity for random state distributed
with measure induced by the superfidelity is greater than the mean purity for
random state distributed with Hilbert-Schmidt distribution.

\section{Generating random states}\label{sec:generating}

\subsection{One qubit case}
In the case of $2 \times 2$ matrices the density function on the simplex of
eigenvalues reads
\begin{equation} \label{eqn:pdf-G2}
f_{\mathrm{G},2} (\lambda,1-\lambda) =  \frac{2 \sqrt{2}}{\pi }  \frac{1}{\sqrt{\lambda (1-\lambda)}}.
\end{equation}

Then the cumulative probability function for eigenvalues by
integrating $f_{G,2}$ over interval $[0,t]$ reads
\begin{equation}
F_{\mathrm{G},2}(t) = \frac{2}{\pi}
\left(\sqrt{(1-t) t}-2 \sqrt{(1-t) t^3}+\arcsin \sqrt{t} \right)
\end{equation}

From the above we obtain a simple method for generating matrices with the above
distribution. First one must generate eigenvalues of the matrix by inverting the
cumulative distribution function and then rotate it by a random unitary matrix
distributed with respect to Haar measure.



\subsection{General case}
To generate random state of dimension $N>2$ distributed with measure induced by
the superfidelity, one can use the rejection method (see \eg{}
\cite{devroye1986non}).

Probability density function $f_{\mathrm{G},N}$ on a simplex of eigenvalues can
be bounded as
\begin{equation}
 f_{\mathrm{G},N}(\lambda) \leq c f_{\mathrm{B},N}(\lambda), \ \ \forall \lambda \in \Delta\label{eqn:rejection-bound}
\end{equation}
where $f_{\mathrm{B},N}$ is a probability density function generated by Bures
measure~\cite{BZ06} (see Fig.~\ref{fig:bures-density-3})
\begin{equation}
 f_{\mathrm{B},N}(\lambda) = C_N^\mathrm{B} \frac{1}{\sqrt{\lambda_1 \ \dots \lambda_N}} \prod_{i < j} 
 \frac{(\lambda_i - \lambda_j)^2}{\lambda_i + \lambda_j}. 
\end{equation}
Indeed, we have
\begin{eqnarray}
\sup_{\lambda} \frac{f_{\mathrm{G},N}(\lambda)}{f_{\mathrm{B},N}(\lambda)} &=&
\frac{C^\mathrm{G}_N }{C^\mathrm{B}_N} \frac{N^{-N/2} (2/N)^{N (N-1)/2}}{\sqrt{1 - 1/N}}  
\end{eqnarray}
and using the bound for $C^\mathrm{G}_N$ one can take 
\begin{equation}
 c = \frac{ \sqrt{\frac{N^2-N}{N^2+1}} \Gamma(N^2) \pi^{N/2} }{ \prod_{i=1}^{N} \Gamma(i) 2^{N(N-1)/2} \Gamma(N^2/2) N^{N^2/2}}
\end{equation}
as the constant in Eq.~(\ref{eqn:rejection-bound}). 

In order to generate a matrix distributed according to the measure induced by
the superfidelity, one needs to generate a random matrix $X$ distributed with
Bures measure~\cite{osipov10randombures} and a random number $u$ distributed
uniformly over the unit interval $[0,1]$. To accept $X$ as a matrix distributed
according to the measure induced by the superfidelity, we check if $u\leq
\frac{1}{c}\frac{f_{\mathrm{G},N}(X)}{f_{\mathrm{B},N}(X)}$ holds.
Unfortunately, constant $c$ increases very rapidly with $N$ and thus this method
does not work very efficiently for large~$N$.

\section{Summary}\label{sec:summary}
We have analysed random density matrices distributed according to probability
measure induced by superfildelity. We have derived the formula for the
probability density of eigenvalues according to this measure. We have also shown
that random states distributed according to this measure have mean purity larger
than in the case of Hilbert-Schmidt measure. We also provide a method for
generating random matrices according to the introduced distribution.

Still there are some problems which require further investigations. The first is
the calculation of the exact formula for the normalization constant for the
probability density function. This is directly related to the distribution of
purity for measures induced by the partial trace~\cite{giraud2007purity,
giraud_distribution_2007}. The second problem is the inefficient method of
sampling random states with the introduced measure, which could be used for 
numerical studies of the geometry of quantum states~\cite{BZ06,dunkl11shadow1,dunkl11shadow2}.

\begin{acknowledgments}
Work of Z.~Pucha{\l}a was partially supported by the Polish National Science
Centre under the research project N N514 513340 and partially by the Polish
Ministry of Science and Higher Education under the research project IP 2010 033
470. Work of J.~A.~Miszczak was partially supported by the Polish National
Science Centre under the research project N N516 475440 and partially by the
Polish Ministry of Science and Higher Education under the research project IP
2010 052 270.  

Authors would like to thank K.~\.Zyczkowski and P.~Gawron for motivation and
interesting discussions.
\end{acknowledgments}

\bibliographystyle{apsrev4-1}
\bibliography{super-volume}

\begin{thebibliography}{10}%
\makeatletter
\providecommand \@ifxundefined [1]{%
 \ifx #1\undefined \expandafter \@firstoftwo
 \else \expandafter \@secondoftwo
\fi
}%
\providecommand \@ifnum [1]{%
 \ifnum #1\expandafter \@firstoftwo
 \else \expandafter \@secondoftwo
\fi
}%
\providecommand \enquote [1]{``#1''}%
\providecommand \bibnamefont  [1]{#1}%
\providecommand \bibfnamefont [1]{#1}%
\providecommand \citenamefont [1]{#1}%
\providecommand\href[0]{\@sanitize\@href}%
\providecommand\@href[1]{\endgroup\@@startlink{#1}\endgroup\@@href}%
\providecommand\@@href[1]{#1\@@endlink}%
\providecommand \@sanitize [0]{\begingroup\catcode`\&12\catcode`\#12\relax}%
\@ifxundefined \pdfoutput {\@firstoftwo}{%
 \@ifnum{\z@=\pdfoutput}{\@firstoftwo}{\@secondoftwo}%
}{%
 \providecommand\@@startlink[1]{\leavevmode\special{html:<a href="#1">}}%
 \providecommand\@@endlink[0]{\special{html:</a>}}%
}{%
 \providecommand\@@startlink[1]{%
  \leavevmode
  \pdfstartlink
   attr{/Border[0 0 1 ]/H/I/C[0 1 1]}%
   user{/Subtype/Link/A<</Type/Action/S/URI/URI(#1)>>}%
  \relax
 }%
 \providecommand\@@endlink[0]{\pdfendlink}%
}%
\providecommand \url  [0]{\begingroup\@sanitize \@url }%
\providecommand \@url [1]{\endgroup\@href {#1}{\urlprefix}}%
\providecommand \urlprefix [0]{URL }%
\providecommand \Eprint[0]{\href }%
\@ifxundefined \urlstyle {%
  \providecommand \doi [1]{doi:\discretionary{}{}{}#1}%
}{%
  \providecommand \doi [0]{doi:\discretionary{}{}{}\begingroup
  \urlstyle{rm}\Url }%
}%
\providecommand \doibase [0]{http://dx.doi.org/}%
\providecommand \Doi[1]{\href{\doibase#1}}%
\providecommand \bibAnnote [3]{%
  \BibitemShut{#1}%
  \begin{quotation}\noindent
    \textsc{Key:}\ #2\\\textsc{Annotation:}\ #3%
  \end{quotation}%
}%
\providecommand \bibAnnoteFile [2]{%
  \IfFileExists{#2}{\bibAnnote {#1} {#2} {\input{#2}}}{}%
}%
\providecommand \typeout [0]{\immediate \write \m@ne }%
\providecommand \selectlanguage [0]{\@gobble}%
\providecommand \bibinfo [0]{\@secondoftwo}%
\providecommand \bibfield [0]{\@secondoftwo}%
\providecommand \translation [1]{[#1]}%
\providecommand \BibitemOpen[0]{}%
\providecommand \bibitemStop [0]{}%
\providecommand \bibitemNoStop [0]{.\EOS\space}%
\providecommand \EOS [0]{\spacefactor3000\relax}%
\providecommand \BibitemShut [1]{\csname bibitem#1\endcsname}%
\bibitem{zyczkowski11generating}%
  \BibitemOpen
  \bibfield{author}{%
  \bibinfo {author} {\bibfnamefont{K.}~\bibnamefont{\.Zyczkowski}}, \bibinfo
  {author} {\bibfnamefont{K.~A.}\ \bibnamefont{Penson}}, \bibinfo {author}
  {\bibfnamefont{I.}~\bibnamefont{Nechita}},\ and\ \bibinfo {author}
  {\bibfnamefont{B.}~\bibnamefont{Collins}},\ }%
  \bibfield{journal}{%
  \Doi{10.1063/1.3595693}{\bibinfo {journal} {J. Math. Phys.}}\ }%
  \textbf{\bibinfo {volume} {52}},\ \bibinfo {pages} {062201} (\bibinfo {year}
  {2011}),\ \Eprint{http://arxiv.org/abs/1010.3570}{arXiv:1010.3570}%
  \bibAnnoteFile{NoStop}{zyczkowski11generating}%
\bibitem{BZ06}%
  \BibitemOpen
  \bibfield{author}{%
  \bibinfo {author} {\bibfnamefont{I.}~\bibnamefont{Bengtsson}}\ and\ \bibinfo
  {author} {\bibfnamefont{K.}~\bibnamefont{{\.Z}yczkowski}},\ }%
  \emph{\bibinfo {title} {{Geometry of Quantum States: An Introduction to
  Quantum Entanglement}}}\ (\bibinfo {publisher} {Cambridge University Press,
  Cambridge, U.K.},\ \bibinfo {year} {2006})%
  \bibAnnoteFile{NoStop}{BZ06}%
\bibitem{Uh76}%
  \BibitemOpen
  \bibfield{author}{%
  \bibinfo {author} {\bibfnamefont{A.}~\bibnamefont{Uhlmann}},\ }%
  \bibfield{journal}{%
  \Doi{10.1016/0034-4877(76)90060-4}{\bibinfo {journal} {Rep. Math. Phys.}}\ }%
  \textbf{\bibinfo {volume} {9}},\ \bibinfo {pages} {273} (\bibinfo {year}
  {1976})%
  \bibAnnoteFile{NoStop}{Uh76}%
\bibitem{caves}%
  \BibitemOpen
  \bibfield{author}{%
  \bibinfo {author} {\bibfnamefont{C.~M.}\ \bibnamefont{Caves}},\ }%
  \enquote{\bibinfo {title} {{Measures and volumes for spheres, the probability
  simplex, projective Hilbert space, and density operators}},}\  (\bibinfo
  {year} {2001}),\ \bibinfo {note}
  {\url{http://info.phys.unm.edu/~caves/reports/measures.pdf}}%
  \bibAnnoteFile{NoStop}{caves}%
\bibitem{subsuper}%
  \BibitemOpen
  \bibfield{author}{%
  \bibinfo {author} {\bibfnamefont{J.~A.}\ \bibnamefont{Miszczak}}, \bibinfo
  {author} {\bibfnamefont{Z.}~\bibnamefont{Pucha{\l}a}}, \bibinfo {author}
  {\bibfnamefont{P.}~\bibnamefont{Horodecki}}, \bibinfo {author}
  {\bibfnamefont{A.}~\bibnamefont{Uhlmann}},\ and\ \bibinfo {author}
  {\bibfnamefont{K.}~\bibnamefont{\.Zyczkowski}},\ }%
  \bibfield{journal}{%
  \bibinfo {journal} {Quantum Inform. Comput.}\ }%
  \textbf{\bibinfo {volume} {9}},\ \bibinfo {pages} {0103} (\bibinfo {year}
  {2009}),\ \Eprint{http://arxiv.org/abs/0805.2037}{arXiv:0805.2037}%
  \bibAnnoteFile{NoStop}{subsuper}%
\bibitem{Hubner92}%
  \BibitemOpen
  \bibfield{author}{%
  \bibinfo {author} {\bibfnamefont{M.}~\bibnamefont{H{\"u}bner}},\ }%
  \bibfield{journal}{%
  \Doi{10.1016/0375-9601(92)91004-B}{\bibinfo {journal} {Phys. Lett. A.}}\ }%
  \textbf{\bibinfo {volume} {163}},\ \bibinfo {pages} {239} (\bibinfo {year}
  {1992})%
  \bibAnnoteFile{NoStop}{Hubner92}%
\bibitem{Hall98}%
  \BibitemOpen
  \bibfield{author}{%
  \bibinfo {author} {\bibfnamefont{M.~J.~W.}\ \bibnamefont{Hall}},\ }%
  \bibfield{journal}{%
  \Doi{10.1016/S0375-9601(98)00190-X}{\bibinfo {journal} {Phys. Lett. A.}}\ }%
  \textbf{\bibinfo {volume} {242}},\ \bibinfo {pages} {123} (\bibinfo {year}
  {1998}),\
  \Eprint{http://arxiv.org/abs/quant-ph/9802052}{arXiv:quant-ph/9802052}%
  \bibAnnoteFile{NoStop}{Hall98}%
\bibitem{buresvolume}%
  \BibitemOpen
  \bibfield{author}{%
  \bibinfo {author} {\bibfnamefont{K.}~\bibnamefont{\.Zyczkowski}}\ and\
  \bibinfo {author} {\bibfnamefont{H.-J.}\ \bibnamefont{Sommers}},\ }%
  \bibfield{journal}{%
  \Doi{10.1088/0305-4470/36/39/308}{\bibinfo {journal} {J. Phys. A: Math.
  Gen.}}\ }%
  \textbf{\bibinfo {volume} {36}},\ \bibinfo {pages} {10083} (\bibinfo {year}
  {2003}),\
  \Eprint{http://arxiv.org/abs/quant-ph/0304041}{arXiv:quant-ph/0304041}%
  \bibAnnoteFile{NoStop}{buresvolume}%
\bibitem{sommers04statistical}%
  \BibitemOpen
  \bibfield{author}{%
  \bibinfo {author} {\bibfnamefont{H.-J.}\ \bibnamefont{Sommers}}\ and\
  \bibinfo {author} {\bibfnamefont{K.}~\bibnamefont{\.Zyczkowski}},\ }%
  \bibfield{journal}{%
  \Doi{10.1088/0305-4470/37/35/004}{\bibinfo {journal} {J. Phys. A: Math.
  Gen.}}\ }%
  \textbf{\bibinfo {volume} {37}},\ \bibinfo {pages} {8457} (\bibinfo {year}
  {2004}),\
  \Eprint{http://arxiv.org/abs/quant-ph/0405031}{arXiv:quant-ph/0405031}%
  \bibAnnoteFile{NoStop}{sommers04statistical}%
\bibitem{mendonca}%
  \BibitemOpen
  \bibfield{author}{%
  \bibinfo {author} {\bibfnamefont{P.~E. M.~F.}\ \bibnamefont{Mendonca}},
  \bibinfo {author} {\bibfnamefont{R.}~\bibnamefont{d.~J.~Napolitano}},
  \bibinfo {author} {\bibfnamefont{M.~A.}\ \bibnamefont{Marchiolli}}, \bibinfo
  {author} {\bibfnamefont{C.~J.}\ \bibnamefont{Foster}},\ and\ \bibinfo
  {author} {\bibfnamefont{Y.-C.}\ \bibnamefont{Liang}},\ }%
  \bibfield{journal}{%
  \Doi{10.1103/PhysRevA.78.052330}{\bibinfo {journal} {Phys. Rev. A}}\ }%
  \textbf{\bibinfo {volume} {78}},\ \bibinfo {pages} {052330} (\bibinfo {year}
  {2008}),\ \Eprint{http://arxiv.org/abs/0806.1150}{arXiv:0806.1150}%
  \bibAnnoteFile{NoStop}{mendonca}%
\bibitem{process-superfidelity}%
  \BibitemOpen
  \bibfield{author}{%
  \bibinfo {author} {\bibfnamefont{Z.}~\bibnamefont{Pucha{\l}a}}, \bibinfo
  {author} {\bibfnamefont{J.~A.}\ \bibnamefont{Miszczak}}, \bibinfo {author}
  {\bibfnamefont{P.}~\bibnamefont{Gawron}},\ and\ \bibinfo {author}
  {\bibfnamefont{B.}~\bibnamefont{Gardas}},\ }%
  \bibfield{journal}{%
  \Doi{10.1007/s11128-010-0166-1}{\bibinfo {journal} {Quant. Inf. Proc.}}\ }%
  \textbf{\bibinfo {volume} {10}},\ \bibinfo {pages} {1} (\bibinfo {year}
  {2011}),\ \Eprint{http://arxiv.org/abs/0911.0567}{arXiv:0911.0567}%
  \bibAnnoteFile{NoStop}{process-superfidelity}%
\bibitem{boundtrace}%
  \BibitemOpen
  \bibfield{author}{%
  \bibinfo {author} {\bibfnamefont{Z.}~\bibnamefont{Pucha{\l}a}}\ and\ \bibinfo
  {author} {\bibfnamefont{J.~A.}\ \bibnamefont{Miszczak}},\ }%
  \bibfield{journal}{%
  \Doi{10.1103/PhysRevA.79.024302}{\bibinfo {journal} {Phys. Rev. A}}\ }%
  \textbf{\bibinfo {volume} {79}},\ \bibinfo {pages} {024302} (\bibinfo {year}
  {2009}),\ \Eprint{http://arxiv.org/abs/0811.2323}{arXiv:0811.2323}%
  \bibAnnoteFile{NoStop}{boundtrace}%
\bibitem{englert96fringe}%
  \BibitemOpen
  \bibfield{author}{%
  \bibinfo {author} {\bibfnamefont{B.-G.}\ \bibnamefont{Englert}},\ }%
  \bibfield{journal}{%
  \Doi{10.1103/PhysRevLett.77.2154}{\bibinfo {journal} {Phys. Rev. Lett.}}\ }%
  \textbf{\bibinfo {volume} {77}},\ \bibinfo {pages} {2154} (\bibinfo {month}
  {Sep}\ \bibinfo {year} {1996})%
  \bibAnnoteFile{NoStop}{englert96fringe}%
\bibitem{chen2010super}%
  \BibitemOpen
  \bibfield{author}{%
  \bibinfo {author} {\bibfnamefont{Z.}~\bibnamefont{Chen}}, \bibinfo {author}
  {\bibfnamefont{Z.}~\bibnamefont{Ma}}, \bibinfo {author}
  {\bibfnamefont{F.}~\bibnamefont{Zhang}},\ and\ \bibinfo {author}
  {\bibfnamefont{J.}~\bibnamefont{Chen}},\ }%
  \bibfield{journal}{%
  \Doi{10.2478/s11534-010-0123-8}{\bibinfo {journal} {Cent. Eur. J. Phys.}},\
  \bibinfo {pages} {1}}%
   (\bibinfo {year} {2011}),\
  \Eprint{http://arxiv.org/abs/0811.3453}{arXiv:0811.3453}%
  \bibAnnoteFile{NoStop}{chen2010super}%
\bibitem{giraud_distribution_2007}%
  \BibitemOpen
  \bibfield{author}{%
  \bibinfo {author} {\bibfnamefont{O.}~\bibnamefont{Giraud}},\ }%
  \bibfield{journal}{%
  \Doi{10.1088/1751-8113/40/11/014}{\bibinfo {journal} {J. Phys. A: Math.
  Theor.}}\ }%
  \textbf{\bibinfo {volume} {40}},\ \bibinfo {pages} {2793} (\bibinfo {year}
  {2007})%
  \bibAnnoteFile{NoStop}{giraud_distribution_2007}%
\bibitem{giraud2007purity}%
  \BibitemOpen
  \bibfield{author}{%
  \bibinfo {author} {\bibfnamefont{O.}~\bibnamefont{Giraud}},\ }%
  \bibfield{journal}{%
  \Doi{10.1088/1751-8113/40/49/F03}{\bibinfo {journal} {J. Phys. A: Math.
  Theor.}}\ }%
  \textbf{\bibinfo {volume} {40}},\ \bibinfo {pages} {F1053} (\bibinfo {year}
  {2007})%
  \bibAnnoteFile{NoStop}{giraud2007purity}%
\bibitem{scott2003entangling}%
  \BibitemOpen
  \bibfield{author}{%
  \bibinfo {author} {\bibfnamefont{A.~J.}\ \bibnamefont{Scott}}\ and\ \bibinfo
  {author} {\bibfnamefont{C.~M.}\ \bibnamefont{Caves}},\ }%
  \bibfield{journal}{%
  \Doi{10.1088/0305-4470/36/36/308}{\bibinfo {journal} {J. Phys. A: Math.
  Gen.}}\ }%
  \textbf{\bibinfo {volume} {36}},\ \bibinfo {pages} {9553} (\bibinfo {year}
  {2003}),\
  \Eprint{http://arxiv.org/abs/quant-ph/0305046}{arXiv:quant-ph/0305046}%
  \bibAnnoteFile{NoStop}{scott2003entangling}%
\bibitem{tong1997relationship}%
  \BibitemOpen
  \bibfield{author}{%
  \bibinfo {author} {\bibfnamefont{Y.~L.}\ \bibnamefont{Tong}},\ }%
  \bibfield{journal}{%
  \Doi{doi:10.1155/S1025583497000064}{\bibinfo {journal} {J. Inequal. Appl.}}\
  }%
  \textbf{\bibinfo {volume} {1}},\ \bibinfo {pages} {85} (\bibinfo {year}
  {1997})%
  \bibAnnoteFile{NoStop}{tong1997relationship}%
\bibitem{devroye1986non}%
  \BibitemOpen
  \bibfield{author}{%
  \bibinfo {author} {\bibfnamefont{L.}~\bibnamefont{Devroye}},\ }%
  \emph{\bibinfo {title} {{Non-uniform random variate generation}}}\ (\bibinfo
  {publisher} {Springer-Verlag, New York, U.S.A.},\ \bibinfo {year} {1986})%
  \bibAnnoteFile{NoStop}{devroye1986non}%
\bibitem{osipov10randombures}%
  \BibitemOpen
  \bibfield{author}{%
  \bibinfo {author} {\bibfnamefont{V.~A.}\ \bibnamefont{Osipov}}, \bibinfo
  {author} {\bibfnamefont{H.-J.}\ \bibnamefont{Sommers}},\ and\ \bibinfo
  {author} {\bibfnamefont{K.}~\bibnamefont{\.Zyczkowski}},\ }%
  \bibfield{journal}{%
  \Doi{10.1088/1751-8113/43/5/055302}{\bibinfo {journal} {J. Phys. A: Math.
  Theor.}}\ }%
  \textbf{\bibinfo {volume} {43}},\ \bibinfo {pages} {055302} (\bibinfo {year}
  {2010}),\ \Eprint{http://arxiv.org/abs/0909.5094}{arXiv:0909.5094}%
  \bibAnnoteFile{NoStop}{osipov10randombures}%
\bibitem{dunkl11shadow1}%
  \BibitemOpen
  \bibfield{author}{%
  \bibinfo {author} {\bibfnamefont{C.~F.}\ \bibnamefont{Dunkl}}, \bibinfo
  {author} {\bibfnamefont{P.}~\bibnamefont{Gawron}}, \bibinfo {author}
  {\bibfnamefont{J.~A.}\ \bibnamefont{Holbrook}}, \bibinfo {author}
  {\bibfnamefont{Z.}~\bibnamefont{Pucha{\l}a}},\ and\ \bibinfo {author}
  {\bibfnamefont{K.}~\bibnamefont{\.Zyczkowski}},\ }%
  \bibfield{journal}{%
  \Doi{10.1016/j.laa.2010.12.003}{\bibinfo {journal} {Linear Alg. Appl.}}\ }%
  \textbf{\bibinfo {volume} {434}},\ \bibinfo {pages} {2042} (\bibinfo {year}
  {2011}),\ \Eprint{http://arxiv.org/abs/1010.4189}{arXiv:1010.4189}%
  \bibAnnoteFile{NoStop}{dunkl11shadow1}%
\bibitem{dunkl11shadow2}%
  \BibitemOpen
  \bibfield{author}{%
  \bibinfo {author} {\bibfnamefont{C.}~\bibnamefont{Dunkl}}, \bibinfo {author}
  {\bibfnamefont{P.}~\bibnamefont{Gawron}}, \bibinfo {author}
  {\bibfnamefont{J.~A.}\ \bibnamefont{Holbrook}}, \bibinfo {author}
  {\bibfnamefont{J.~A.}\ \bibnamefont{Miszczak}}, \bibinfo {author}
  {\bibfnamefont{Z.}~\bibnamefont{Pucha{\l}a}},\ and\ \bibinfo {author}
  {\bibfnamefont{K.}~\bibnamefont{\.Zyczkowski}},\ }%
  \bibfield{journal}{%
  \bibinfo {journal} {J. Phys. A: Math. Theor.}\ }%
  \bibinfo {note} {in press},\
  \Eprint{http://arxiv.org/abs/1104.2760}{arXiv:1104.2760}%
  \bibAnnoteFile{NoStop}{dunkl11shadow2}%
\end{thebibliography}%

\end{document}